\documentclass[9pt, conference]{IEEEtran}
\IEEEoverridecommandlockouts
\usepackage{amsmath,amssymb,amsfonts}
\usepackage{capt-of}% or \usepackage{caption}
\usepackage{algorithmic}
\usepackage{graphicx}
\usepackage{textcomp}
\usepackage{xcolor}
\usepackage{quoting,xparse}
\usepackage[accsupp]{axessibility}
\usepackage{mathtools}
\usepackage{booktabs}
\usepackage{tipa}
\usepackage{comment}
\usepackage[pagebackref,breaklinks,colorlinks]{hyperref}

\def\BibTeX{{\rm B\kern-.05em{\sc i\kern-.025em b}\kern-.08em
    T\kern-.1667em\lower.7ex\hbox{E}\kern-.125emX}}

\begin{document}

\title{Speech2rtMRI: Speech-Guided Diffusion Model for Real-time MRI Video of the Vocal Tract during Speech\\
% {\footnotesize \textsuperscript{*}Note: Sub-titles are not captured in Xplore and
% should not be used}
% \thanks{This work was supported by the National Science Foundation (NSF) under Grant No. IIS 2311676. The content is solely the responsibility of the authors and does not necessarily represent the official views of the NSF.}
}
\author{
  \IEEEauthorblockN{
    Hong Nguyen\IEEEauthorrefmark{2},\quad
    Sean Foley\IEEEauthorrefmark{2},\quad
    Kevin Huang\IEEEauthorrefmark{2},\quad
    Xuan Shi\IEEEauthorrefmark{2},\quad
    Tiantian Feng\IEEEauthorrefmark{2},\quad
    Shrikanth Narayanan\IEEEauthorrefmark{2}
  }
  \IEEEauthorblockA{
    \IEEEauthorrefmark{2}Signal Analysis and Interpretation Lab,
    University of Southern California, Los Angeles, CA 90089
  }
  % \IEEEauthorblockA{
  %   \IEEEauthorrefmark{2}Roski Eye Institute,
  %   University of Southern California, Los Angeles, CA 90089
  % }
}

\maketitle

\begin{abstract}
Understanding speech production both visually and kinematically can inform second language learning system designs, as well as the creation of speaking characters in video games and animations. 
In this work, we introduce a data-driven method to visually represent articulator motion in Magnetic Resonance Imaging (MRI) videos of the human vocal tract during speech based on arbitrary audio or speech input. We leverage large pre-trained speech models, which are embedded with prior knowledge, to generalize the visual domain to unseen data using an speech-to-video diffusion model. Our findings demonstrate that the visual generation significantly benefits from the pre-trained speech representations.
We also observed that evaluating phonemes in isolation is challenging but becomes more straightforward when assessed within the context of spoken words.
Limitations of the current results include the presence of unsmooth tongue motion and video distortion when the tongue contacts the palate. The source code is available for the public at: \href{https://github.com/Hong7Cong/SPAN-rtmri.git}{https://github.com/Hong7Cong/SPAN-rtmri.git}
\end{abstract}


\section{Introduction}
% \begin{pquotation}{Richard Feynman, 1918}
% `What I cannot create, I do not understand'
% \end{pquotation} 
Humans produce speech by directing air from the lungs through the vocal tract, with this airflow then being shaped by the movements of the articulators to create the desired sounds. Decades of experimental work suggest this process is also modulated by higher-level linguistic representations, such as phonemes or articulatory gestures \cite{goldstein2007dynamic, guenther2016neural,parrell2019current} (Fig. 1). % [cite]. 
 A common approach to understanding speech production is via an acoustic-to-articulatory inversion (AAI) study. AAI methods estimate articulatory movements from an acoustic input to serve a wide range of applications:
\begin{itemize}
    \item Animation: Speech-driven animation \cite{Animation06, prabhune2023towards} of the face and inner mouth in interactive systems or multimedia like video games
    \item Language learning: Visual aids assist in learning second language (L2) pronunciation \cite{Suemitsu2015ARA, Levitt2010TheEO, Gick200811UI} and help hearing-impaired individuals with speech acquisition.
    \item Synthetic data generation/simulation: Create synthetic data through inversion such as for testing scientific hypotheses including in speech pathology and diseases affecting speech production \cite{Hagedorn2018EngineeringInnovationinSpeech,hagedorn2021complexity}.
\end{itemize}

Previous works on AAI involve both statistical mapping and neural deep learning approaches. Several models from both approaches have been proposed to predict the vocal tract movement in terms of electromagnetic articulography (EMA) features, including but not limited to the use of Gaussian Mixture models \cite{Ghosh2013Onsmoothingarticulatorytrajectories}, Deep Convolution Networks \cite{Csapo2020SpeakerDA, Illa2019RepresentationLU}, Recurrent Networks \cite{csapo2020speaker, Illa2019AnIO}, and Generative adversarial network \cite{CiwaGAN24, avatarcontrolnature23, prabhune2023towards}. Other works \cite{Bigioi2023SpeechDV} investigate AAI via synthesized lip movement using audio-conditioned GAN and Diffusion models.
% These approaches created several challenges, such as errors of EMA location across individuals, and require prior experts' knowledge of EMA.
These works often rely on a point-tracking articulatory database (EMA, X-ray Microbeam) or lip movements, which only reflect a partial capture of the vocal tract, while the entire vocal tract airway can be captured via real-time MRI \cite{Narayanan2004Anapproachtoreal-time}. The rtMRI modality, for example, allows for visualization of the complete midsagittal plane, including the jaw, lips, larynx, velum, and tongue.
Using rtMRI, Tamas \cite{csapo2020speaker} proposed LSTM models for direct acoustic-to-articulatory inversion. However, this work only considered four subjects for training and did not take into account cross-speaker vocal tract dynamics.
More recently, Sathvik \cite{Udupa2022RealTimeMV} proposed a phoneme-conditioned variational autoencoder (CVAE). The work requires extraction and alignment of specific phonemes before training the CVAE model.
These models were also trained on a limited set of speakers, and the acquired knowledge remains constrained within the boundaries of the dataset.
Hence, there is room to further develop novel AAI approaches that can generalize better, leveraging the increasing availability of rtMRI speech production data \cite{Lim2021AMD}.
% In contrast, infants do not need any linguistic background beforehand to mimic sounds. 

\begin{figure}[t]
\centering
\centerline{\includegraphics[width=\linewidth]{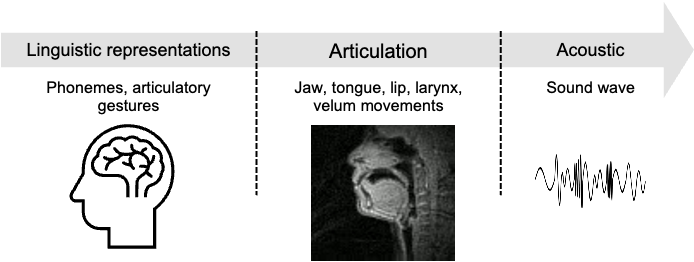}}
   % \caption{Overview of our speech-2-rtMRI Diffusion modeling framework for generating vocal tract movement video during speech. Our modeling framework includes two main phases: training and sampling.}
  \caption{The speech chain from higher-level linguistic representations to acoustic output. Our focus in this work is on the low-level articulation with the aim to generate vocal tract movements conditioned on acoustic prompts.} %Factors related to speech perception and co, like semantic meaning, syntactic or emotion of speech, are not investigated in this work.}
\label{fig:exampleofeyetract}
\end{figure}

\begin{figure*}[t]
  \centering
   \includegraphics[width=\linewidth]{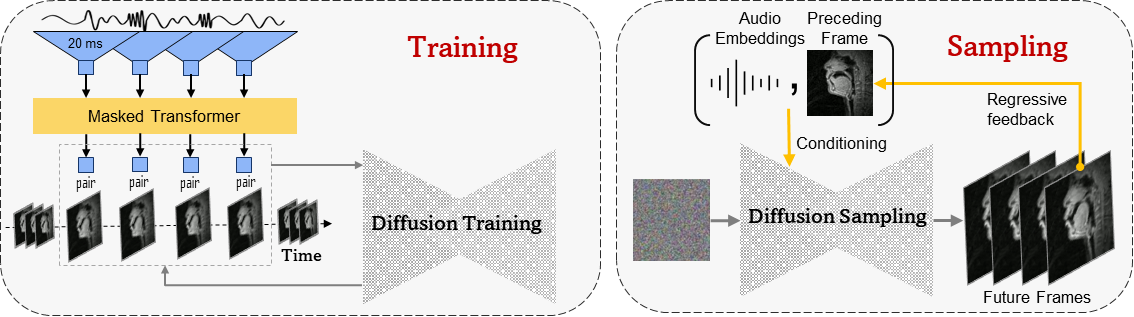}
   \caption{Overview of our speech-2-rtMRI Diffusion modeling framework for generating vocal tract movement video during speech. Our modeling framework includes two main phases: training and sampling.}
   \label{fig:system}
\end{figure*}

With recent progress in generative diffusion models and large audio models, it has become possible to synthesize speech production visually via the movements of articulators while take advantage of pre-trained acoustic knowledge. Like infants acquiring speech by refining their vocal tract articulation to produce sounds they've heard, the diffusion process iteratively refines gaussian noise to create images/videos.
% This work proposes an audio-guided video diffusion model for generating magnetic resonance videos of articulators. 
The fundamental challenge in developing this video generation model is the inherent variability in human speech, which is influenced by the plurality of factors such as accent, age, and gender.
We leverage pre-trained speech models as prior knowledge to support synthesizing articulatory motion in the visual domain. 

In this paper, we train an audio-to-video diffusion model using raw audio (primarily speech) and a corresponding video of articulator motion, both recorded simultaneously in an MRI machine. The model learns joint distribution between %freezer
audio embeddings and articulator visual space to synthesize real-time MRI (rtMRI) videos for any given speech audio.
Our contributions in this paper are summarized as follows
\begin{itemize}
    \item We propose Speech2rtMRI, a Speech-Conditioned Diffusion model to synthesize real-time MRI video of articulator movement during speech. 
    \item We conduct objective evaluations on the synthesized articulatory movements to assess the effectiveness of our approach. Results show that the WavLM model, particularly the large version, achieves the best overall generalization scores.
    \item In addition to objective metrics, we include human evaluations involving phoneticians to rate the relevance, authenticity and limitations of the generated real-time MRI videos. The human evaluation results also highlight that the generated videos contain frequent unsmooth vocal tract movements, with the tongue, in particular, displaying unnatural motion. 
\end{itemize}

% The challenge also the opportunities of diffusion in medical domain is data sharpness [cite] which defined as 
\begin{table*}[t]
\caption{Evaluation results of a data-driven audio-to-video diffusion approach across different pre-trained speech models, using the metrics FVD (Fréchet Video Distance) and SSIM (Structural Similarity Index Measure) on an unseen test set.}
% \newlength\mylength
% \setlength\mylength{\dimexpr.5\columnwidth-2\tabcolsep-0.5\arrayrulewidth\relax}
\centering
\setlength{\tabcolsep}{8pt}
\begin{tabular}{l c c c c c c c c}\toprule
&& & \multicolumn{3}{c}{Unseen (FVD) $\downarrow$} & \multicolumn{3}{c}{Unseen (SSIM) $\uparrow$}\\
\cmidrule(r){4-6} \cmidrule(r){7-9}  
Audio Model & Size & Pooling & Speech  & Subject & Both & Speech  & Subject & Both\\\midrule
$\mathrlap{\textit{Real Video in test set (Upper Bound)}}$  \\
Real Video &-    &-  & 34 & 70 & 80 & 25.8 & 26.2 & 26.7\\
% &$\checkmark$  &&& - & -& 7.1 & 32023 & 1.1e-11 & 1.4e1\\
% &$\checkmark$    &&& - & 8.0e-9 & 4.3 & 17348 & 1.5e-11 & 6.6 \\
$\mathrlap{\textit{Our Approach: Audio-to-Video Diffusion}}$  \\
HuBERT & Base &$\times$  & 1541 $\pm$ 20 & 1571 $\pm$ 20 & 1594 $\pm$ 20 & 10.32 $\pm$ 0.19 & 10.99 $\pm$ 0.21 & 11.06 $\pm$ 0.24  \\
 & Large &$\times$  & 1540 $\pm$ 19 & 1581 $\pm$ 19 & 1616 $\pm$ 19 & 10.79 $\pm$ 0.18 & 11.36 $\pm$ 0.22 & 11.43 $\pm$ 0.25\\
  & Large &$\checkmark$  & 1425 $\pm$ 21 & 1478 $\pm$ 22 & 1499 $\pm$ 21 & \textbf{11.20 $\pm$ 0.19} & \textbf{11.63 $\pm$ 0.20} & {11.71 $\pm$ 0.21} \\
WavLM & Base  &$\times$ & 1475 $\pm$ 18 & 1523 $\pm$ 19 & 1529 $\pm$ 18 & 10.72 $\pm$ 0.21 & 11.46 $\pm$ 0.23 & 11.49 $\pm$ 0.20  \\
& Large  &$\times$ & 1426 $\pm$ 18 & 1479 $\pm$ 18 & \textbf{1473 $\pm$ 17} & 10.72 $\pm$ 0.19& 11.32 $\pm$ 0.21& 11.41 $\pm$ 0.25  \\
 & Large &$\checkmark$  & \textbf{1420 $\pm$ 19} & \textbf{1466 $\pm$ 19} & 1493 $\pm$ 19 & {11.03 $\pm$ 0.20} & {11.61 $\pm$ 0.22} & \textbf{11.71 $\pm$ 0.25}   \\
Wav2Vec2 & Base & $\times$ & 1757 $\pm$ 25 & 1764 $\pm$ 25 & 1759 $\pm$ 25 & 6.95 $\pm$ 0.10 & 7.25 $\pm$ 0.11 & 7.26 $\pm$ 0.13  \\
& Large  & $\times$ & 1590 $\pm$ 23 & 1639 $\pm$ 24 & 1651 $\pm$ 24 & 10.87 $\pm$ 0.18 & 11.31 $\pm$ 0.19 & 11.39 $\pm$ 0.20  \\
 % & Large  &$\times$ & 1553 & - & - & - & - & - &-&-&-&-\\
 & Large  &$\checkmark$ & 1596 $\pm$ 24 & 1642 $\pm$ 24 & 1637 $\pm$ 24 & 10.37 $\pm$ 0.18 & 10.86 $\pm$ 0.19 & 10.92 $\pm$ 0.21  \\
\bottomrule
\end{tabular}\label{table:baselineresults}
\end{table*}

\section{Speech-to-rtMRI Diffusion Model}
\subsection{Overview}
Our approach consists of two primary stages, similar to conventional diffusion models: training and sampling, as shown in Fig.~\ref{fig:system}. In the training stage, we input sequences of frames into a 3D Diffusion U-Net to learn joint distribution between pre-trained audio space and target visual spaces. The 3D Diffusion U-Net takes three inputs: the original sequences of frames along with video metadata (batch size, frames, color channels, height, and width), the corresponding audio embeddings for each sequence, and the preceding frames representing the vocal tract at rest. The specific function of each input will be explained in the following subsection. 
Once the training is complete, diffusion uses a sampling technique to generate synthetic video using arbitrary audio embedding and an initial frame. The output samples have the same length of training videos. We propose to use the Regressive Feedback Technique to generate longer videos, such as for word production.

\subsection{Training}
The pre-trained speech model uses self-supervised learning to learn speech representations from large amounts of unlabeled audio data in the time domain.
Specifically, in this work, we compare several widely recognized pre-trained speech models as the speech feature encoder, including Wave2Vec2 \cite{baevski2020wav2vec}, HuBERT \cite{hsu2021hubert}, and WavLM \cite{chen2022wavlm}.

% For a fair comparison, we utilize the Wav2Vec-based family of models, including Wave2Vec2 [cite], Hubert [cite], and WavLM [cite], as our speech feature encoder.

% Due to the nature of the problems, we do not consider mel-spectrum speech models such as Whisper because we want to focus on audio in the time domain.

\noindent
\textbf{Speech Conditioning} For given sequences of video frames, there is a corresponding sequence of speech embeddings. The speech embeddings are derived from the segmented speech signals aligned with the corresponding video frames. 
We encode the whole speech sentence and distribute sequences of embeddings to pair with sequences of frames, rather than encoding each 200 ms segment of raw audio independently. As a result, each embedding carries contextual information about the entire sentence. 
% This ensures that the speech embeddings retain contextual knowledge of the entire speech signal.
% We extract speech embeddings from the segmented speech signal pair with the sequence of video frames. This approach ensures that the speech embeddings retain contextual knowledge of the entire speech signal.
% akin to the planning phase \cite{martin2010planning} of speech production in the human brain.
This mechanism of speech conditioning used in the training is classifier-free guidance (CFG). 
%In CFG setting, 

\noindent
\textbf{Initial Frame Conditioning} We train the diffusion model conditioned on both speech embeddings and the initial frame. The first frame guides the video generation to align with the predefined vocal tract shape. The mechanism involved in Initial Frame Conditioning is in-painting, in which 3D-Unet starts sampling from the concatenation of the initial frame and pure Gaussian noise.

\noindent
\textbf{Pooling} Training our diffusion model involves implementing an attention mechanism between speech embeddings and the sequence of frames. For simplicity, we do pooling by averaging a sequence of speech embeddings. 
The possible advantage of pooling lies in its ability to average out noise within the embeddings; however, this process may also inadvertently eliminate temporal information that can be important for the generation. 
To answer this, we specifically explore whether a simple averaging of the speech embeddings within the attention module will provide advantages for training the speech-to-video diffusion model.

\subsection{Sampling}

\noindent
\textbf{Regressive Feedback with Preceding Frame.} At inference time, we sample a long videos with arbitrary length by stacking short generated videos. We condition the model with the preceding frame and the speech embedding corresponding to its timeframe. There are two main reasons behind this. One is to save computation efficiency for generative models to learn long videos by training on short videos and combining them. Secondly, because the fundamental structure of a sentence involves the combination of words and words themselves are composed of combinations of phonemes, it would be more intuitive to learn from low-level phonetic information and infer higher-level structure, e.g., phonemes, syllables, and words. 

\begin{table*}[ht]
\begin{minipage}[ht]{0.58\linewidth}
\centering
\caption{F1-Score \& Mean Opinion Score (MOS) of human evaluation on Speech2rtMRI generation of words. MOS-GT is the score for the ground truth videos, while MOS-A2V is the score for the samples generated by Speech-to-Video.}
\begin{tabular}{lccccccccc}
\toprule
\multicolumn{1}{c}{} & \multicolumn{8}{c}{\textbf{Word (Small sample size)}} \\
\cmidrule(rl){2-9}
\textbf{} & {bite} & {bat} & {beat} & {bird} & {bit} & {boat} & {bought} & {butte}\\
\midrule
F1      & 0.68 & 0.68 & 0.67 & 0.68 & 0.67 & 0.68 & 0.69 & 0.68 \\
MOS-GT  & 3.27 & 3.19 & 2.75 & 2.04 & 3.64 & 3.00 & 3.20 & 2.54 \\
MOS-A2V & 1.21 & 1.46 & 1.21 & 1.26 & 1.39 & 1.57 & 1.21 & 1.19 \\
\bottomrule
\end{tabular}
\label{table:student}
\end{minipage}\hfill
\begin{minipage}[ht]{0.42\linewidth}
\centering
\includegraphics[width=0.3\linewidth] {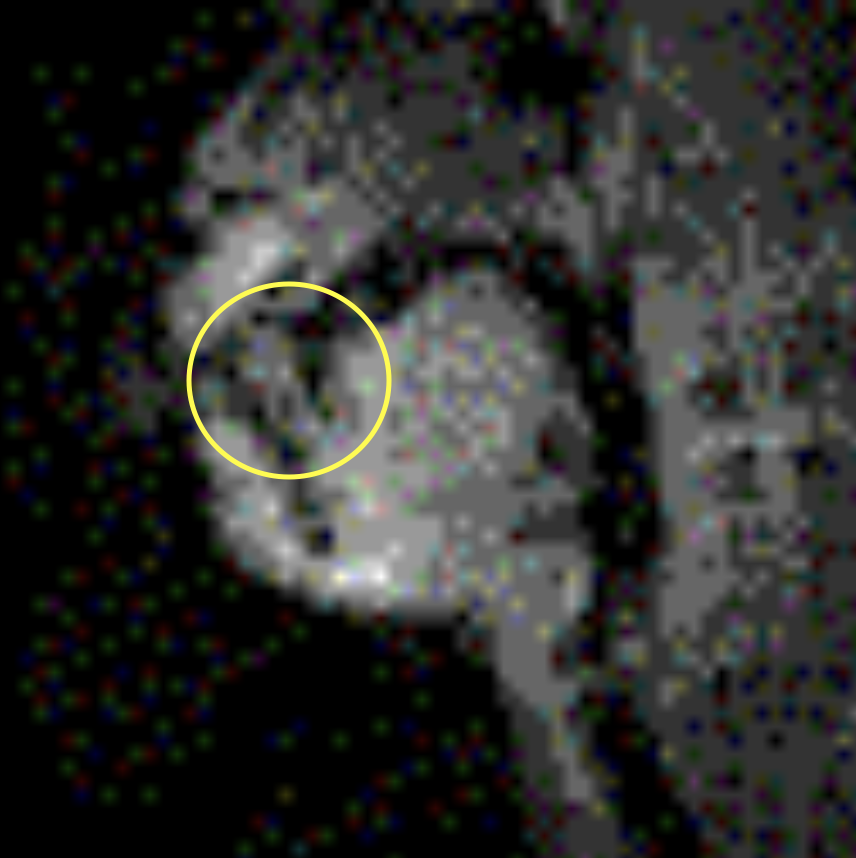}
\includegraphics[width=0.3\linewidth] {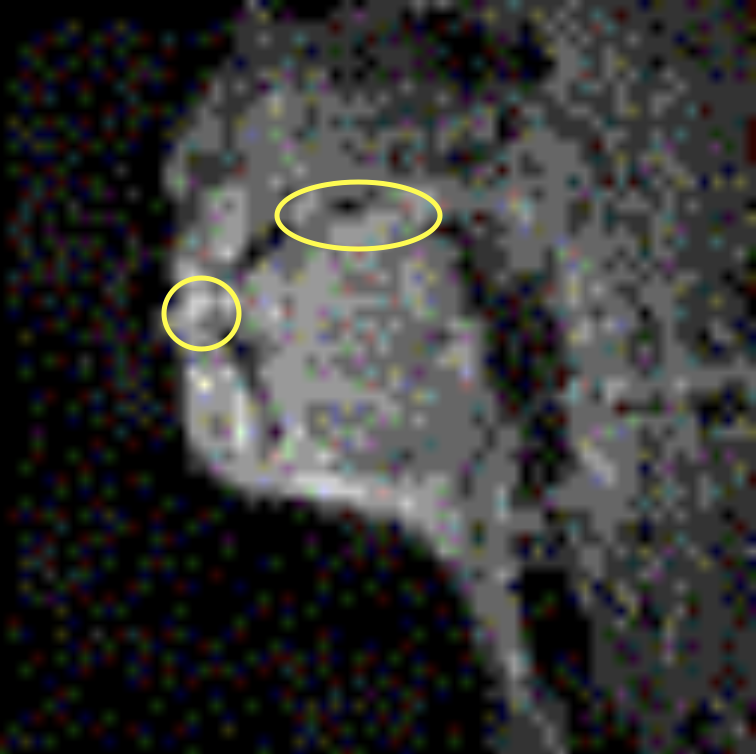}
\includegraphics[width=0.3\linewidth] {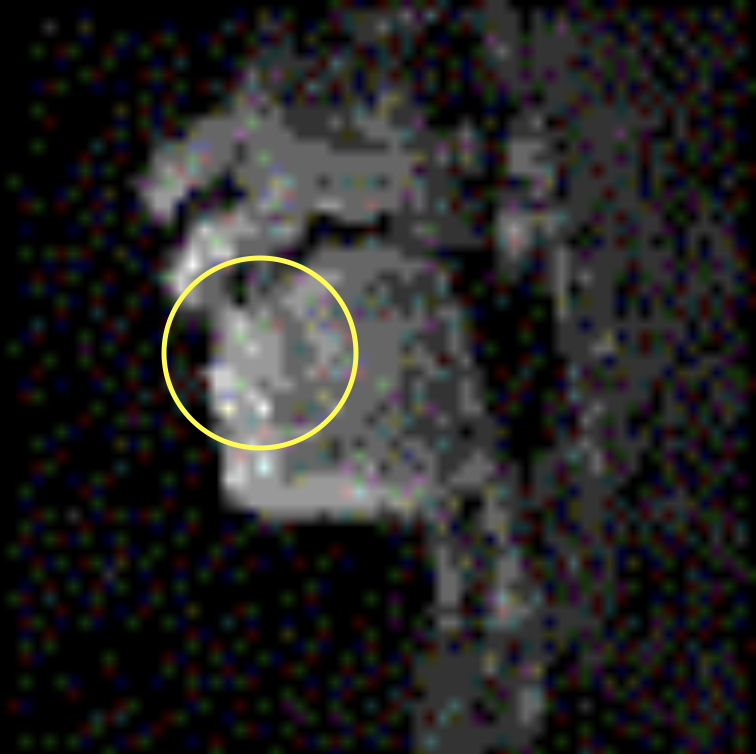}
\captionof{figure}{Example cases of video quality degradation during generation. Left image show inauthentic tongue shapes while middle and right images show points of tongue-palate contact before quality degradation.}
\label{fig:samplesrtmri}
\end{minipage}
\end{table*}

% \begin{table}
% \caption{Average Accuracy \& scores of human evaluation on audio-to-word generation}
% \centering
% \begin{tabular}{lccccccccc}
% \toprule
% \multicolumn{1}{c}{} & \multicolumn{8}{c}{\textbf{Word (Small sample size)}} \\
% \cmidrule(rl){2-9}
% \textbf{} & {bite} & {bat} & {beat} & {bird} & {bit} & {boat} & {bought} & {butte}\\
% \midrule
% Accuracy & 0.79 & 0.76 & 0.74 & 0.75 & 0.75 & 0.74 & 0.78 & 0.75 \\
% Score & 2.5 & 2.4 & 2.2 & 1.9 & 2.4 & 2.4 & 2.4 & 2.2 \\
% % CommonVoice & - & - & - & - & - \\
% % 300-hours Switchboard & - & - & - & - & - \\
% \bottomrule
% \end{tabular}
% \end{table}

%\hfill
%\parbox{.60\linewidth}{
%\caption{Accuracy of human evaluation on audio-to-words generation}
%\centering
%\begin{tabular}{lccccccccc}
%\toprule
%\multicolumn{1}{c}{} & \multicolumn{8}{c}{\textbf{Word (Small sample size)}} \\
%\cmidrule(rl){2-9}
%\textbf{} & {bite} & {bat} & {beat} & {bird} & {bit} & {boat} & {bought} & {butte}\\
%\midrule
%Phoneticians 1 & 00.0 & 00.0 & 00.0 & 00.0 & 00.0 & 00.0 & 00.0 & 00.0 \\
%Phoneticians 2 & - & - & - & - & - & - & - & - \\
% CommonVoice & - & - & - & - & - \\
% 300-hours Switchboard & - & - & - & - & - \\
%\bottomrule
%\end{tabular}
%}
%\end{table*}

\section{Experiments}
\subsection{Dataset}
We conduct our experiments on the 75-Speaker USC Speech MRI dataset \cite{Lim2021AMD}. This dataset provides a unique collection of 2D midsagittal-view rtMRI videos paired with synchronized audio from 75 subjects performing linguistically driven speech tasks. Additionally, it features a 3D volumetric MRI of the vocal tract during sustained speech sounds and a high-resolution static anatomical T2-weighted MRI of the upper airway for each subject. The dataset is freely accessible to the research community.

Out of the 75 speakers, we randomly selected 15 speakers for the unseen subject test set. Additionally, for each of the remaining 60 speakers, we select 4 audio recordings on free-form topics to form the unseen audio test set. The remaining audio-video recordings, consisting of fixed-script audio, were used for training the diffusion model.

\subsection{Experiment Settings}
% We used records of 60 subjects for training and 15 for inference. 
Most of the selected pre-trained speech models take speech input at a rate of 16kHz and produce the speech embeddings every 20 milliseconds. The resulting speech embeddings used for conditioning have a size of 768 in all base models and 1024 in all large models.
For the video input, frames are down-sampled to a resolution of 64x64 pixels at 50 frames per second to match the time resolution (20 milliseconds) of audio condition embeddings.
Most pre-trained speech models used in this work are pre-trained with the English speech data.

In the diffusion model training setup, we simply utilized a single 3D-UNet for the generation without considering an additional UNet to enhance the resolution of the output videos. The 3D-UNet takes three inputs in the modeling: ten frames of rtMRI, the corresponding speech embeddings, and the initial reference frame. We perform the diffusion model training on a GTX 3090 with a batch size 2. 

\subsection{Evaluation Metrics}\label{AA}
To quantitatively evaluate our models, we use the widely used Fréchet Video Distance (FVD) and Structural Similarity Index Measure (SSIM) as objective evaluation metrics. For subjective evaluation, we asked trained phoneticians to identify the generated rt-MRI videos within a mixed batch of synthetic and real sets and to rate the generated videos' realism on a scale of 1 to 5. While the FVD measures the realism of generated videos compared to true distribution, the phoneme prediction task evaluates the accuracy of synthesized rt-MRI in capturing specific phoneme production. 

\noindent
\textbf{Fréchet Video Distance.} Motivated by a related concept utilized to assess the quality of generated images \cite{Unterthiner2019FVDAN}, Fréchet Video Distance additionally captures the ``perceptual similarity" of the motion within a video. From \cite{Unterthiner2019FVDAN}, FVD metric is sensitive not only to visual degradation but also to the motion of video frames. 

\noindent
\textbf{Structural Similarity Index Measure} is an image-level metric used \cite{wang2004image} for evaluating video quality. It is a full-reference visual quality assessment index based on brightness, contrast, and structure.

\noindent
\textbf{Human Evaluation.} FVD only shows how real the video is in motion but does not reflect the reliability of generated rt-MRI on a specific phoneme or word. To evaluate the perceived quality of the generative model guided by audio features, we hired two well-trained phoneticians who are familiar with rtMRI imaging for speech production to perform the following two tasks:
\begin{itemize}
    \item Real-or-Synthesized Classification: We uniformly mixed synthesized and real videos of people producing a set of words, then gave them to phoneticians to identify whether each sample is real or not. We gave some real samples for each word beforehand to phoneticians for reference.
    \item For each word sample that is identified as ``not real," phoneticians gave a score on a scale from 1-5 to indicate how natural the synthesized sample is. The higher is better. At the end of each word section, phoneticians gave qualitative comments on what generally makes the generated samples look fake.
\end{itemize}speech models, includinuBERT, WaveLM, and Wav2Vec2 videoh are used as the conditioning models for gener{atile~\ref{table:baselineresults}}using diffusion modeling training. The evaluation is based on unseen data in three scenarios: ``Speech," ``Subject," and ``Both," using two key metrics: Fréchet Video Distance (FVD) and Structural Similarity Index Measure (SSIM).
Overall, the results demonstrate that synthesizing rtMRI for unseen subjects is more challenging than synthesizing unseen speech, possibly due to the large variations in vocal tract shape and pronouncing pattern across different speakers.
On the other hand, the conditioning using speech embeddings of HuBERT and WavLM yields similar performances in video generation, especially in their ``Large" versions, where they achieved strong results in both FVD and SSIM metrics. 
Specifically, we notice that most models produce promising results compared to real videos in generating synthetic videos measured by structural similarity. 
% This finding is more noticeable when the test condition is unseen speech. 
We want to highlight that real videos represent the upper-bound performance for the generation task and are inherently associated with the highest FVD and SSIM scores. 

\subsection{Motion fidelity is as important as visual fidelity in medical domain} 
Synthetic video realism can be categorized into visual realism and motion realism. In medical applications, where the quality of videos or images may be affected by noise, preserving motion fidelity is just as important as achieving visual realism. Although both FVD and SSIM measure the visual fidelity of target samples, FVD is better at capturing the authenticity of motion \cite{liufrechet}. As shown in Table \ref{table:baselineresults}, while SSIM results show no significant differences across different pre-trained speech embeddings, FVD results highlight a clear trend where using WavLM as the conditioning yields better generations than other pre-trained speech models. 
% This indicates that pre-trained speech features enhance the generalization of motion fidelity but contribute less to visual realism. 

% \subsection{Synthetic phonemes are harder to assess than words}  
% We made attempt to evaluate synthetic phonemes but not working due to the following reason.
% At the articulatory level, phonemes are not produced sequentially, but rather overlap considerably. Thus, accurate generation would entail accounting for these coarticulatory effects, wherein the execution of a given phoneme is highly dependent on the surrounding context. For example, in a typical consonant-vowel sequence, such as [ti] in ``teen", the raising of the tongue tip for /t/ and the raising of the tongue body for /i/ are produced simultaneously \cite{fowler1980coarticulation}. A viable model would generate this ``coproduction" effect, rather than generating such movements sequentially. Our diffusion model utilizes a fixed training length of 200 ms, which aligns with the average duration of articulatory motion for most vowels. For consonants, the average duration is even smaller, around 20 ms, much smaller than the duration between audio embeddings and video frames. 

\subsection{Human evaluation} 
As seen in Table \ref{table:student}, the F1 scores of the human evaluators averaged around 0.68 for most words, while the MOS of the generated samples were in the lower range, roughly around 1.2-1.6. The consistency of the F1 scores and MOS across words suggests that the model is fairly consistent in its generations across different vowels, which is important for applications in medical domains. Despite the generated samples having lower MOS, it should be noted that the ground-truth samples did not score particularly high either, ranging from 2.0-3.6. This may be attributed to the evaluation having both binary classification and Likert scale tasks, wherein an evaluator is likely to rate it lower if they already classified it as generated. In terms of qualitative analysis, both evaluators noted instances of the vocal tract movements not being smooth, with the tongue in particular exhibiting unnatural movements, as shown in Fig.~\ref{fig:samplesrtmri}. This finding is expected, given the variability of movements exhibited by the tongue during speech production. Furthermore, Fig.~\ref{fig:samplesrtmri} demonstrates that the generated video quality deteriorates significantly when the tongue makes contact with the hard palate and alveolar ridge. This may be attributed to the speed and precision of movements involving the tongue tip or oral constrictions more generally making distinctions between active and passive articulators less clear. 
% This issue may be attributed to an imbalance in the dataset used to train the diffusion model, possibly due to the lack of videos with the tongue when it is not in contact with other articulators. Addressing this problem in future work may involve increasing the training dataset size or balancing the dataset more effectively.

%Note that the word with the lowest average score being ``bird" is unsurprising given the range of articulatory postures used to produce rhotics in American English \cite{delattre1968dialect}. The training data likely exhibits such variability, making it difficult for the model to learn a stable pattern for the vowel in this word. 

\section{Discussion}
\subsection{Synthetic phonemes are harder to assess than words}
We attempted to evaluate the synthetic phonemes; however, this was unsuccessful for the following reasons.
At the articulatory level, phonemes are not produced sequentially, but rather overlap considerably. Thus, accurate generation would entail accounting for these coarticulatory effects, wherein the execution of a given phoneme is highly dependent on the surrounding context. For example, in a typical consonant-vowel sequence, such as [ti] in ``teen", the raising of the tongue tip for /t/ and the raising of the tongue body for /i/ are produced simultaneously \cite{fowler1980coarticulation}. A viable model would generate this ``coproduction" effect, rather than generating such movements sequentially. Our diffusion model utilizes a fixed training length of 200 ms, which aligns with the average duration of articulatory motion for most vowels. For consonants, the average duration is even smaller, around 20 ms, much smaller than the duration between audio embeddings and video frames. 

\subsection{Automatic Synthetic phoneme/word/sentence evaluation metric} 
Current metrics, such as FVD and SSIM, primarily assess the realism of motion and visual quality of the generated content without considering the accuracy of the sound-articulator relationship. Moreover, relying on linguistic experts to evaluate audio-to-articulator outputs for every model is neither practical nor cost-effective. Therefore, new evaluation metrics are needed that can effectively assess the generation of articulatory videos during the production of phonemes, words, and sentences. 

% \subsection{Variability of speech production} 
% Due to the articulator compensation, a variety of articulatory movements may generate identical pronunciation, which breaks the one-to-one relationship between articulatory and acoustic features. Many attributes affect the variation of producing phonemes/words, including context and emotion.

\subsection{Differential weighting of the articulators} Future work should likely aim to focus on generating accurate movements of the tongue, given that during speech production this is the organ that undergoes the largest range of movement, making it also the most challenging aspect of the video the generate accurately. Unlike other articulators in the vocal tract constrained by attachments to bone, the tongue is a hydrostat (a muscle with no skeletal support) \cite{kier1985tongues}, allowing it greater freedom of movement.

As with the human acquisition of speech motor learning, in which control over the lips and jaw are mastered before the tongue \cite{namasivayam2020speech}, our models here also showed a better generation of jaw and lip movements compared to the tongue qualitatively. 
Additionally, accurate tongue movements are essential for producing most speech sounds, while inaccurate jaw and lip movements may have less impact on the overall effectiveness of the generated video of articulatory motion. 

\subsection{Training data diversity} 
One notable behavior is that the quality of long-generated videos tends to degrade over time. This decline may be attributed to the regressive feedback mechanism, where the last frame of a sequence is used to condition the first frame of the subsequent sequence. 
The last frame of every sample degrades from true distribution till the images are completely destroyed.

\section{Conclusion}
We proposed a speech-to-video diffusion model for synthesizing articulatory movements during speech. Our findings suggest that the generative model generalizes better to unseen speech samples than unseen subjects. We also note that the proposed generative framework yields the best generation scores when using WavLM speech embeddings as the condition. In the human evaluation experiment, linguistic experts were able to accurately identify synthetic videos with a recall of 1, although they frequently misclassified true negative samples. Experts recommend giving more attention to realistic tongue movement to enhance the quality of the generated videos, such as dealing with tongue-palate contact events. 

\bibliographystyle{unsrt}
\bibliography{refs}

\end{document}